# Rapid Determination of Antimicrobial Susceptibility by Stimulated Raman Scattering Imaging of D$_2$O Metabolic Incorporation in a Single Bacterium

Meng Zhang[1], Nader S. Abutaleb[2], Junjie Li[1], Pu-Ting Dong[1], Cheng Zong[1], Pu Wang[3,4], Mohamed N. Seleem[2,5], Weili Hong[3,]*, Ji-Xin Cheng[1,6,]*

[1] Department of Electrical & Computer Engineering, Boston University, Boston, Massachusetts 02215, USA.
[2] Department of Comparative Pathobiology, Purdue University, West Lafayette, Indiana 47907, USA.
[3] Beijing Advanced Innovation Center for Biomedical Engineering, Beihang University, Beijing, 100191, China
[4] Vibronix Inc., West Lafayette, Indiana 47906, USA.
[5] Purdue Institute of Inflammation, Immunology, and Infectious Disease, West Lafayette, Indiana 47907, USA.
[6] Department of Biomedical Engineering, Department of Chemistry, Photonics Center, Boston University, Boston, Massachusetts 02215, USA.
*Corresponding authors: jxcheng@bu.edu; weilihong@buaa.edu.cn.

**Abstract**

Rapid antimicrobial susceptibility testing (AST) is urgently needed for treating infections with correct antibiotics and slowing down the emergence of antibiotic-resistant bacteria. Here, we report a phenotypic platform that rapidly produces AST results by femtosecond stimulated Raman scattering imaging of deuterium oxide ($D_2O$) metabolism. Metabolic incorporation of $D_2O$ into biomass in a single bacterium and the susceptibility to antibiotics are probed in as short as 10 minutes after culture in 70% $D_2O$ medium, the fastest among current technologies. Single-cell metabolism inactivation concentration (SC-MIC) is obtained in less than 2.5 h from colony to results. The SC-MIC results of 37 sets of samples, which include 8 major bacterial species and 14 different antibiotics often encountered in clinic, are validated by standard minimal inhibitory concentration blindly measured via broth microdilution. Towards clinical translation, SRS imaging of $D_2O$ metabolic incorporation and SC-MIC determination after 1-h antibiotics treatment and 30-minutes mixture of $D_2O$ and antibiotics incubation of bacteria in urine or whole blood is demonstrated.

**1. Introduction**

Antimicrobial resistance has become a growing public threat, causing nearly 1 million related mortality each year globally.[1] It was estimated that by 2050, antimicrobial resistance will cause 10 million deaths and $100 trillion global production loss if no action is taken.[1-2] To combat this crisis, rapid antimicrobial susceptibility testing (AST) is essential to slow down the emergence of antimicrobial resistance and consequently reduce the deaths caused by drug-resistant infections.[3] The gold standard for AST is conducted by disc diffusion or broth dilution methods and used to determine whether the bacteria are susceptible, intermediate or resistant to antimicrobial agents tested.[4] After 16 to 20-h growth, an minimal inhibitory concentration (MIC) value is read as complete growth inhibition through visual inspection. Current culture-based phenotypic method for AST is too slow to guide immediate decision for infectious disease

treatment.[5] For clinical samples, it usually takes at least 24 h for bacterial preincubation and at least additional 16 h for AST.[6] Genotypic methods[7] do not rely on culturing and provide faster results, but they only target specific known genetic sequences with resistance and thus, are not generally applicable to different bacterial species or mechanisms of resistance, nor providing MIC results.[8]

To overcome these limitations, novel phenotypic methods for rapid AST are under development,[9] including microfluidic devices that increase the detection sensitivity by confining the sample in a small area,[10] monitoring bacterial growth or morphological changes at single cell level,[10e, 11] phenotypic AST quantifying the nucleic acids copy number,[8a, 12] and Raman spectroscopy that probes the chemical content inside a bacterium.[13] While these methods reduce the time for AST, most of them only work for bacterial isolates.

Inside a cell, NADPH is ubiquitously used for biomolecular synthesis.[14] Based on rapid enzyme-catalyzed exchange between the redox-active H atom in NADPH and the D atom in deuterium oxide ($D_2O$), so called heavy water, cellular metabolic activity can be probed via monitoring the intracellular conversion of $D_2O$ into C-D bonds of the biomolecules. Spontaneous Raman spectroscopy of metabolic incorporation of $D_2O$ has been used to determine antimicrobial susceptibility.[15] However, the small spontaneous Raman scattering cross section does not allow high-throughput AST. By spontaneous Raman measurement, it usually takes ca. 10 minutes (30 second per spectrum) to acquire Raman spectra of 20 individual bacteria. Thus, to determine MIC via 10 concentrations of one antibiotics, the total Raman measurment time per strain would be 100 minutes. Thus, it would need at least 17 h to determine MIC of 10 antibiotics for one strain. In contrast, by focusing the excitation energy on the C-D vibration band, coherent Raman microscopy based on either coherent anti-Stokes Raman scattering (CARS) or stimulated Raman scattering (SRS) provides orders-of-magnitude signal enhancement, thereby enabling high-speed and high-throughput chemical imaging of single cells.[16] For broad Raman bands such as CH and CD stretch vibrations, femtosecond pulse SRS further boosts the signal level.[17]

Here, we report a rapid phenotypic platform that can determine the susceptibility of bacteria in urine and whole blood by femtosecond SRS imaging of $D_2O$ metabolism in a single bacterium. Harnessing the high sensitivity of femtosecond SRS imaging, $D_2O$ metabolic incorption inside a signle bacterium to antibiotics is probed in as fast as 10 minutes. Unlike spontaneous Raman spectroscopy, a C-D SRS image covering tens of bacteria is recorded in ~ 1 second. In the presence of antibiotics, a single-cell metabolism inactivation concentration (SC-MIC) is determined in less than 2.5 h from colony to results. Compairsion of SC-MIC results with conventional MIC results among 37 sets of samples, including 8 major bacterial species and 14 different antibiotics often encountered in clinic, yields a category agreement of 94.6% and 5.4% minor error. Moreover, our method is able to determine the metabolic activity and susceptibility of bacteria in either urine or whole blood, which opens the opportunity for rapid single-cell phenotypic AST in clinic.

## 2. Results

**2.1. SRS imaging of $D_2O$ metabolic incorporation in a single bacterium.**

In cells, flavin enzymes catalyze the H-D exchange between water and NADPH's redox active hydrogen in $D_2O$ containing media. The deuterium labeled NADPH mediates fatty acid synthesis reaction with $D_2O$ incorporation, resulting in the deuterated fatty acids production (**Figure 1a**). Biosynthetic pathway of deuterated protein is through introducing deuterium atoms from $D_2O$ into reactions of amino acids (AAs).[14, 18] The schematic of our SRS microscope is shown in **Figure 1b**. In brief, spatially and temporally overlapped pump and Stokes pulses are tuned to match the vibrational frequency of Raman-active modes. The SRS signal appears as an intensity gain in the Stokes beam and an intensity loss in the pump beam, which is extracted through a lock-in amplifier. Stimulated Raman loss is measured, in which most excitation power is in the 1040-nm Stokes beam having a high cellular damage threshold. The carbon-deuterium (C-D) vibrational band, which is spectrally differentiated from endogenous Raman bands, is selectively detected with SRS using either chirped or non-chirped femtosecond laser pulses. Previously,[19] we used chirped femtosecond pulses for hyperspectral SRS imaging of C-D bonds in bacteria. To enhance the detection sensitivity and speed up the imaging process, we applied non-chirping femtosecond pulses and increased the signal to noise ratio by ~5 folds (**Figure S4**). With femtosecond SRS, C-D signals from all bacteria in the field of view could be obtained at a speed of ~1.2 s per image of 200 ×200 pixels, at a pixel dwell time of 30 μs. Therefore, femtosecond SRS imaging enables high-speed, high-throughput study of $D_2O$ incorporation at single bacterium level.

We then examined the toxicity of $D_2O$ to bacterial cells. Unlike mammalian cells, bacteria are much more resistant to $D_2O$ toxicity. Our experiments showed that 70% $D_2O$ concentration did not cause severe growth inhibition (**Figure S1**). Thus, we chose 70% $D_2O$ containing medium to culture bacteria in the following studies. By tuning the Raman shift to C-D vibration at ~2162 $cm^{-1}$, strong signals were observed at individual bacteria after culture in $D_2O$ containing medium for 1 h (**Figure 1c** and **Figure S2**). As a control, no C-D signal was observed for bacteria cultured in normal medium (**Figure 1c** and **Figure S2**). These results were further confirmed by SRS spectra (**Figure 1d**) obtained through temporal tuning of chirped pump and Stokes pulses, and spontaneous Raman spectra (**Figure 1e**), both showing a broad peak (from 2070 to 2250 $cm^{-1}$) at C-D vibration only for bacteria cultured in $D_2O$ containing medium. Therefore, SRS imaging at C-D vibrational region provides a good means to monitor $D_2O$ incorporation in a single bacterium.

To verify metabolic $D_2O$ incorporation in bacteria, we measured the cellular metabolic activity kinetics under different incubation conditions (**Figure S3**). As depicted in the SRS images, the live *P. aeruginosa* cells, cultured in $D_2O$ containing medium at 37 °C, had high metabolic activities and exibited increasingly stronger C-D intensities with increased incubation time. In contrast, neither live *P. aeruginosa* incubated at 4 °C, nor formalin-fixed *P. aeruginosa*, incubated at 37 °C, showed observable C-D signals because of the metabolic activity inhibition. Our findings confirm that C-H bonds are unlikely to undergo abiotic H-D exchange. Instead, cellular metabolic activity directly relates to $D_2O$ incorporation, which is reflected by biochemical transformation of forming C-D bonds in newly synthesized biomolecules.[14, 20]

Next, we investigated whether SRS imaging could resolve the fast $D_2O$ incorporation in biomolecule synthesis spatially and temporally. Time-lapse SRS images (**Figure 1f**) and statistical analysis (**Figure 1g**) showed that the average intensity of C-D signals in *P. aeruginosa* increases with time and saturates at ~2 h. With the enhanced detection sensitivity, C-D signals in individual *P. aeruginosa* can be observed after culture in as short as 10 min, which is shorter than the generation time of *P. aeruginosa* (24 to 27 min).[21] These results showed that the $D_2O$ incorporation of bacteria can be detected by SRS within one cell cycle.

With sub-micron spatial resolution, we further observed the differential distribution of C-D signals in 10-min, 30-min and longer culture time (**Figure 1f**). After 10-min culture, a stronger signal was observed at cell periphery than that at the center of bacteria (**Figure 1f** and **Figure 1h**). In contrast, with 30-min and longer culture times, the signal intensity was stronger at the cell center than that at the cell periphery (**Figure 1f** and **Figure 1i**). These results suggest that $D_2O$ is initially incorporated to synthesize lipids in plasma membrane and then used to synthesize proteins and nucleic acids inside the cell. Collectively, our studies demonstrated that $D_2O$ incorporation can be spatially and temporally monitored by SRS imaging at single bacterium level.

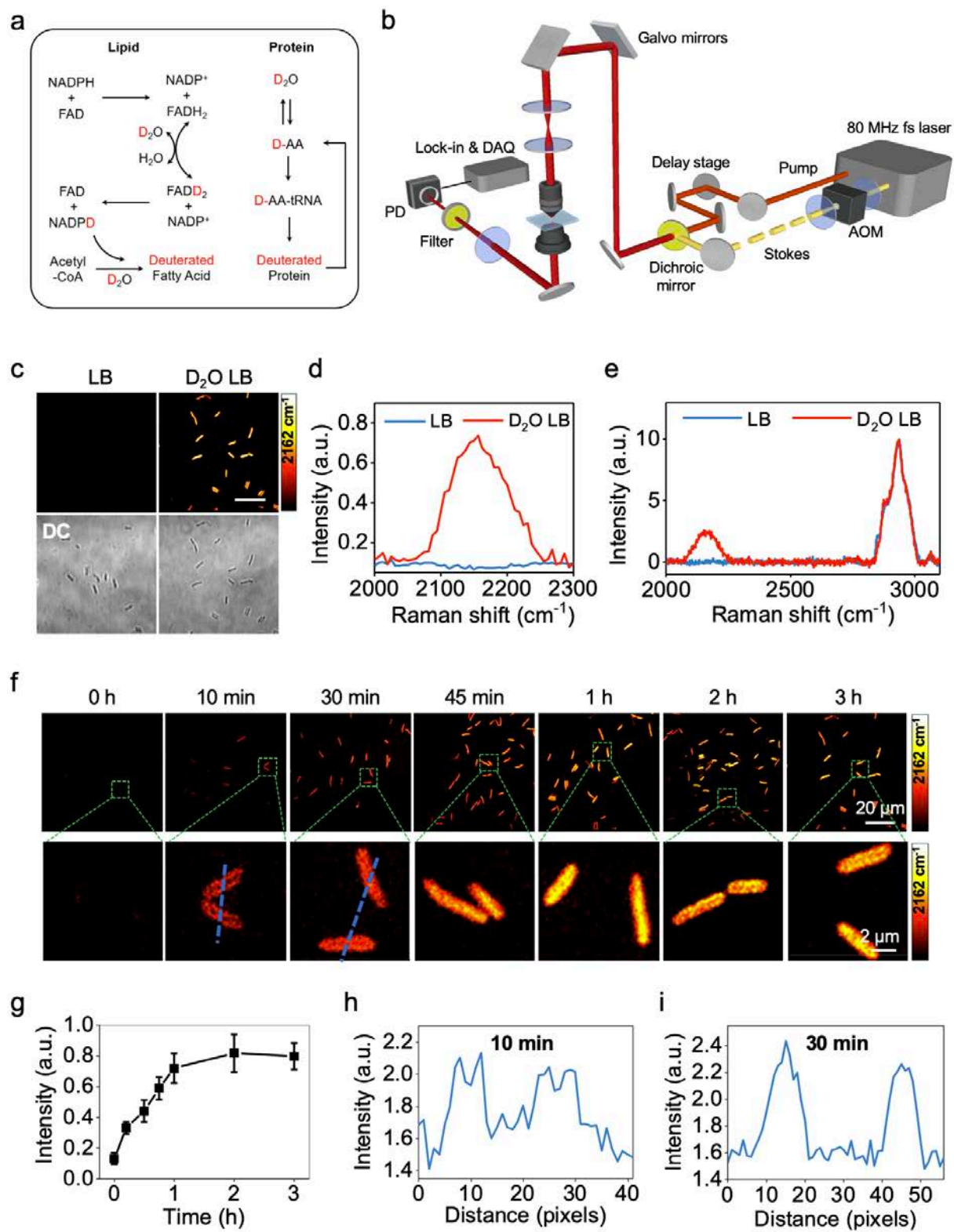

**Figure 1. SRS imaging of D$_2$O metabolic incorporation in a single bacterium.** (a) Scheme for D$_2$O labeling of lipid and protein. (b) Schematic illustration of SRS setup. AOM: acousto-optic modulation. PD: photodiode. Lock-in: lock-in amplifier. DAQ: data acquisition board. (c) SRS and corresponding transmission images of *P. aeruginosa* after culture in normal and D$_2$O-containing medium for 3 h. Scale bar: 20 μm. (d, e) SRS spectra (d) and spontaneous Raman spectra (e) of *P. aeruginosa* after culture in normal and D$_2$O containing medium for 3 h. (f) Time lapse of *P. aeruginosa* after culture in D$_2$O containing medium. (g) Average C-D intensity

plot over time for bacteria in (f). (h, i) Intensity plot of bacteria over the blue line for 10 min culture (h) and 30 min culture (i) in (f). Error bars represent the standard deviation (SD).

## 2.2. D$_2$O incorporation in the presence of antibiotics

To examine how antibiotics affect the metabolic activity of D$_2$O incorporation in bacteria, and to demonstrate that this effect can be used for rapid AST through SRS imaging, a cefotaxime-resistant (MIC = 32 μg/ml) and gentamicin-susceptible (MIC = 4 μg/ml) *P. aeruginosa* strains were selected as a model system. *P. aeruginosa* were cultured for different time in D$_2$O containing medium, with 20 μg/ml gentamicin or cefotaxime. SRS imaging (**Figure 2a**) and statistical analysis (**Figure 2c**) showed that C-D signals of bacteria were significantly reduced after culture with gentamicin, indicating inhibition of D$_2$O incorporation in *P. aeruginosa* by gentamicin. On the contrary, **Figure 2b** and **d** shows that the C-D signals of *P. aeruginosa* cultured with cefotaxime increased with time, indicating active D$_2$O incorporation in bacteria. We also, observed that *P. aeruginosa* tends to form filaments after culture with cefotaxime (**Figure 2b**). This filamentary formation, which happens when Gram-negative bacteria are treated with β-lactam antibiotics, was also observed for *P. aeruginosa* treated with ceftazidime.[10e] Notably, this filamentary formation might be incorrectly interpreted as growth by conventional imaging method.[10e] Yet, it does not affect our metabolic activity measurements.

Next, we examined whether the rapid D$_2$O incorporation inside bacteria can be used to differentiate the antimicrobial susceptibility. We used the relative C-D SRS intensity, the ratio between the antibiotic-treated group and the antibiotic-untreated group (**Figure 1f**), as a biomarker of antimicrobial susceptibility. To determine whether the SRS intensity ratio can be used to distinguish susceptible and resistant groups, the histogram of signal intensities for bacteria after 10-min culture was plotted over the intensity ratio (**Figure 2e**). The plots for susceptible and resistant groups were fitted with normal distribution. A cut-off at 0.62 was determined based on a 10-min culture of bacteria. The large area under curve (AUC = 0.985) in the corresponding receiver operating characteristic (ROC) curve plot clearly demonstrates the ability of this cut-off to separate the two groups (**Figure 2f**). These rusults indicate that our method is capable of determining susceptibility after 10-min $D_2O$ incubation time. The signal intensity ratio between the gentamicin-susceptible and cefotaxime-resistant groups showed more significant difference at longer culture time (**Figure S5**). A cut-off at 0.60 was obtained for the 30-min culture results (**Figure 2g** and **Figure 2h**). In the following studies, we use 30 min of $D_2O$ incubation time to ensure sufficient signal to noise ratio and apply 0.60 cut-off to separate the metabolism active and metabolism inhibited conditions for bacteria cultured at different concentrations of antibiotics. In particular, we use such cut-off to define a single cell metabolism inhibition concentration (SC-MIC) for a certain antibiotic: at or above SC-MIC, the bacteria is sussceptible and thus metabolically inactive; below SC-MIC, the bacteria is resistant and thus metabolically active.

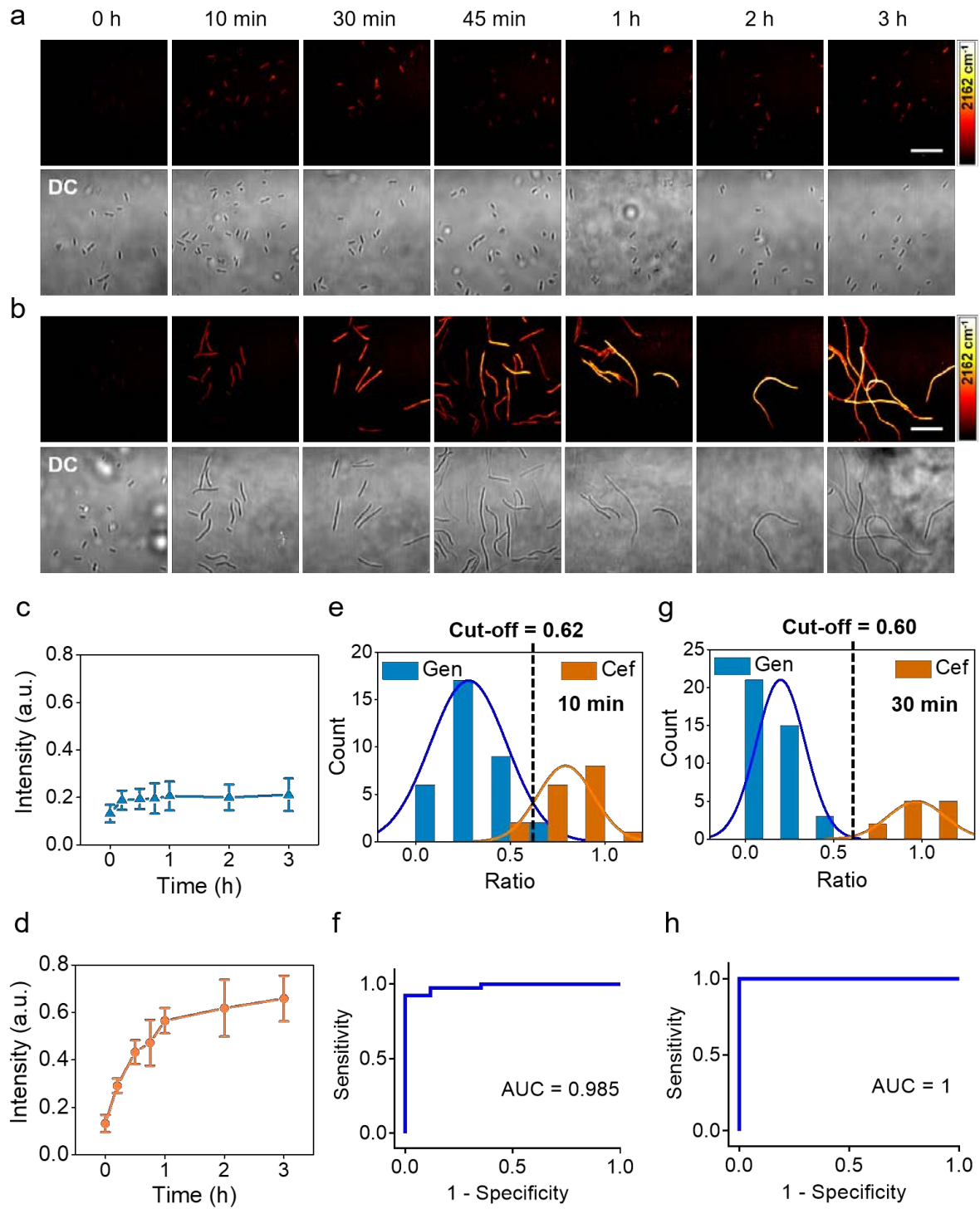

**Figure 2. SRS-based AST of *P. aeruginosa* as a function of culture time.** (a, c) Time lapse SRS at C-D and transmission images of *P. aeruginosa* after culture in $D_2O$-containing medium with the addition of 20 μg/ml gentamicin (a) or cefotaxime (b). (c, d) Average C-D intensity plot over time for *P. aeruginosa* after culture in $D_2O$-containing medium with gentamicin (c) or cefotaxime (d) treatment. Number of cells per group ≥ 10. Error bars represent SD. Scale bar: 20 μm. (e, g) Histogram plot of the count of bacteria as a function of C-D intensity ratio of antibiotic-treated group over the control group after 10 min (e) and 30 min (g) treatment. (f, h) ROC curves of 10 min (f) and 30 min (h) treatment illustrating the ability of the C-D intensity ratio to distinguish susceptible and resistant groups. AUC: area under the curve.

## 2.3. Quantitation of susceptibility via SC-MIC

To explore whether SRS imaging of $D_2O$ metabolic incorporation can quantify the response of bacteria to antibiotics and generate a SC-MIC value comparable with the MIC, we tested *P. aeruginosa* with serially diluted gentamicin. Overnight cultured bacteria were diluted in cation-adjusted Mueller-Hinton Broth (MHB) medium to a final concentration of $8\times10^5$ CFU/ml. The bacteria were first treated with selected antibiotic containing medium for 1 h, then a medium containing $D_2O$ and the same antibiotics was added to bacteria for an additional 30 min (**Figure 3a**). SRS imaging (**Figure 3b**) and statistical analysis (**Figure 3c**) showed that C-D signals at 2 μg/ml or higher gentamicin concentration were significantly lower than that in the control group (0 μg/ml). With the previous determined threshold, $D_2O$ incorporation in *P. aeruginosa* was inhibited at 2 μg/ml and above concentrations. Therefore, the SC-MIC was determined to be 2 μg/ml. This value is within the one-fold difference range with the MIC (4 μg/ml) determined by the broth microdilution method.

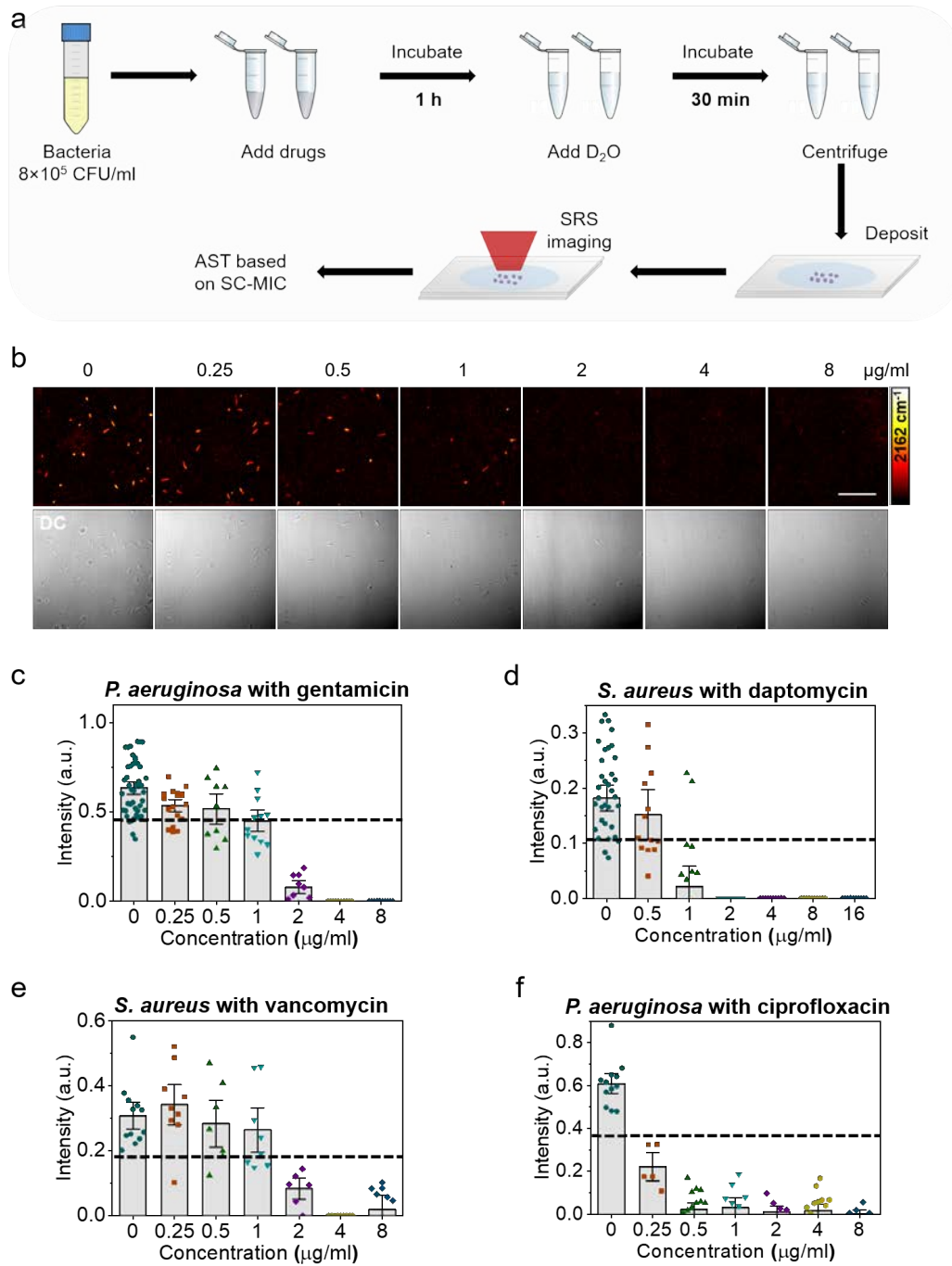

**Figure 3. SC-MIC determination via SRS imaging of $D_2O$ metabolic incorporation in single bacteria.** (a) Workflow of rapid AST with SC-MIC determination by SRS imaging of $D_2O$ metabolic incorporation. (b) SRS at C-D and corresponding transmission images of *P. aeruginosa* after culture in $D_2O$ containing medium with the addition of serially diluted gentamicin. (c) Statistical analysis of C-D intensity in *P. aeruginosa* in (b). (d-f) SC-MIC determination for antibiotics with different mechanisms of action. Error bars represent the standard error of the mean (SEM). Scale bar: 20 μm.

## 2.4. SC-MIC measurement in 37 sets of samples

To validate the broad applicability of our method, we tested 8 major bacterial species and 14 different antibiotics often encountered in clinic (**Table 1**). The antibiotics cover major bacterial inhibition mechanisms of action: inhibition of cell wall synthesis, protein synthesis, DNA synthesis, and/or cell membrane disruption. Typical SRS imaging (**Figure S6**) and statistical analysis (**Figure 3d-3f**) showed that antibiotics with all the mechanisms of action affect $D_2O$ incorporation in bacteria: the β-lactam amoxicillin, the aminoglycoside gentamicin, the fluoroquinolone ciprofloxacin, and the cell membrane targeting daptomycin. We performed 37 sets of the experiments (**Table 1**), where SC-MIC was obtained after 1.0-h incubation with antibiotics and additional 0.5-h incubation with $D_2O$ and antibiotics. For each set, the SC-MIC determination by quantifying the SRS signal intensities versus the concentration of antibiotics is presented as a heatmap. SC-MIC, MIC and the defined susceptibility category interpreted as "susceptible," "resistant," or "intermediate," according to Clinical and Laboratory Standards Institute criteria are presented for each tested bacterial strain. As compared with MIC determined by conventional broth microdilution assay, the SC-MIC (highlighted in black boxes in Table 1) achieved a category agreement of 94.6% (35 out of 37), with 5.4% minor error (2 out of 37), no major error, and no very major error. These results satisfy US Food and Drug Administration (FDA) requirements for AST systems. Most of the SC-MIC results were obtained after 1-h culture in antibiotic containing medium followed by 0.5-h culture in $D_2O$ and antibiotics-containing medium. We observed that methicillin-resistant *S. aureus* (MRSA) grew slower than susceptible *S. aureus*. Therefore, MRSA strains were cultured in $D_2O$ medium for 1 h to achieve comparable C-D signals. With automated imaging and data analysis (**Figure S7**), the whole procedure from colony to results took less than 2.5 h for most of the bacterial strains tested, and 3 h for MRSA strains. Collectively, these results validate SRS imaging of $D_2O$ metabolic incorporation as a rapid and accurate AST method.

We further analyzed the SC-MIC results in the 37 sets of samples based on bacterial species. The 2 minor errors were both from Gram-negative bacteria, resulting in a category agreement of 100% (11 of 11) for Gram-positive samples (9 *S. aureus* samples and 2 *E. faecalis* samples), and a category agreement of 92.3% (24 of 26) with 7.7% minor error (2 of 26) for Gram-negative samples. Though the category agreement in Gram-negative bacterial strains was lower than that in Gram-positive strains, these results still meet the FDA requirements (category agreement ≥ 90%, minor error ≤ 10.0%, major error ≤ 3.0%, very major error ≤ 1.5%).

As shown in **Table 1**, 32 SC-MICs are identical or have one-fold difference with MIC results, resulting in an essential agreement of 86.5% (32 of 37). Four SC-MIC results have three-fold difference, and one result has more than three-fold difference. To better understand the good aggrement and the residual discrepancy between SC-MICs and MICs in these specimens, we obtained MICs of the 37 sets of samples by conventional broth microdilution assay in a blinded manner, using 70% $D_2O$ MHB as the culture medium. The results are listed in **Table S1**. Most of the MICs determined in 70% $D_2O$ MHB are identical or show only one-fold difference with the MICs in normal MHB. Interestingly, for the five results that had the most differences between MIC and SC-MIC, the MICs determined in 70% $D_2O$ MHB agreed more with SC-MICs than MICs determined in normal MHB. Specifically, when *P. aeruginosa* was treated with colistin, a polypeptide that targets bacterial cell membrane, the SC-MIC values were much lower than the MICs in normal MHB, but had much smaller difference from the MICs in 70% $D_2O$ MHB. This comparison indicates that 70% $D_2O$ might increase the venerability of some bacteria to certain antibiotics. The discrepancy between the SC-MICs in 70% $D_2O$ MHB and the MICs in normal MHB can be resolved by using a smaller cut-off value.

**Table 1.** Visualization of SC-MIC results determined from SRS imaging of $D_2O$ metabolic incorporation after 1.5-h incubation time and comparison with MICs determined from gold standard broth microdilution method. The SC-MICs are highlighted with black boxes in the heatmap. *Abbreviations. VAN: vancomycin; LNZ: linezolid; DAP: daptomycin; GEN: gentamicin; ERY: erythromycin; TMP/SMX: trimethoprim/sulfamethoxazole; AMI:

Amikacin; CIP: ciprofloxacin; DOX: doxycycline; TOB: tobramycin; IMI: imipenem; CTX: cefotaxime; AMO: amoxicillin; CL: colistin; S: sensitive; R: resistant; I: Intermediate.

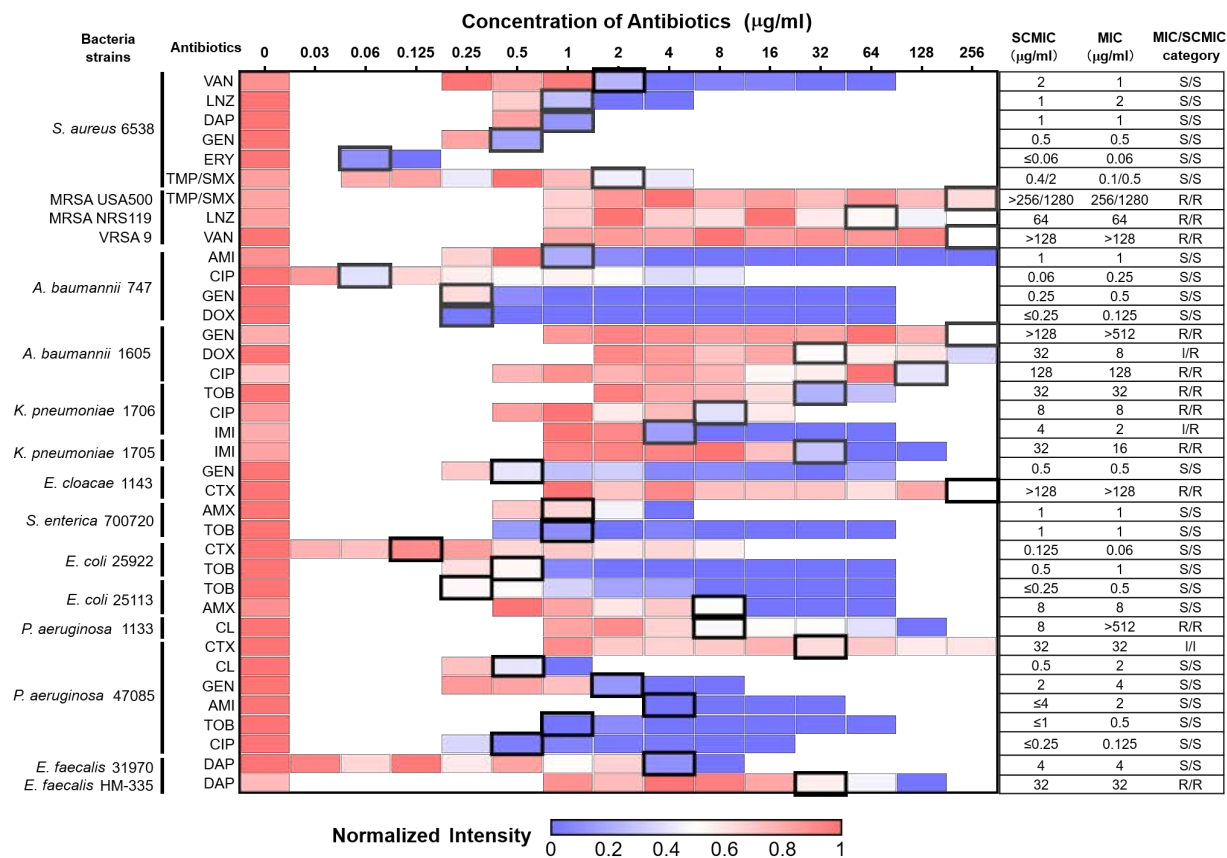

## 2.5. SC-MIC for bacteria in urine environment

To investigate the potential of rapid AST by SRS imaging of $D_2O$ metabolic incorporation for clinical applications, we first tested bacteria in urine. For bacteria in urine, we tested *E. coli*, which is the most common pathogen in urinary tract infection (UTI).[22] To mimic the clinical UTI samples, we used spiked samples by adding *E. coli* to the urine at a final concentration of $10^6$ CFU/ml. After filtration with 5-μm filter and centrifugation, the purified bacteria were used for SC-MIC measurements (**Figure 4a**). This sample preparation procedure took about 15 min. The clean background in SRS images showed that this convenient sample preparation procedure was favorable for rapid AST (**Figure 4b**). SC-MIC for *E. coli* in urine with amoxicillin was determined to be 4 μg/ml (**Figure 4b-4c**), which has the same essential and category agreement with the SC-MIC or MIC for pure *E. coli* (**Figure 4d**). These results showed that rapid AST by

SRS imaging of D$_2$O metabolic incorporation is suitable for clinical application to bacteria in the urine.

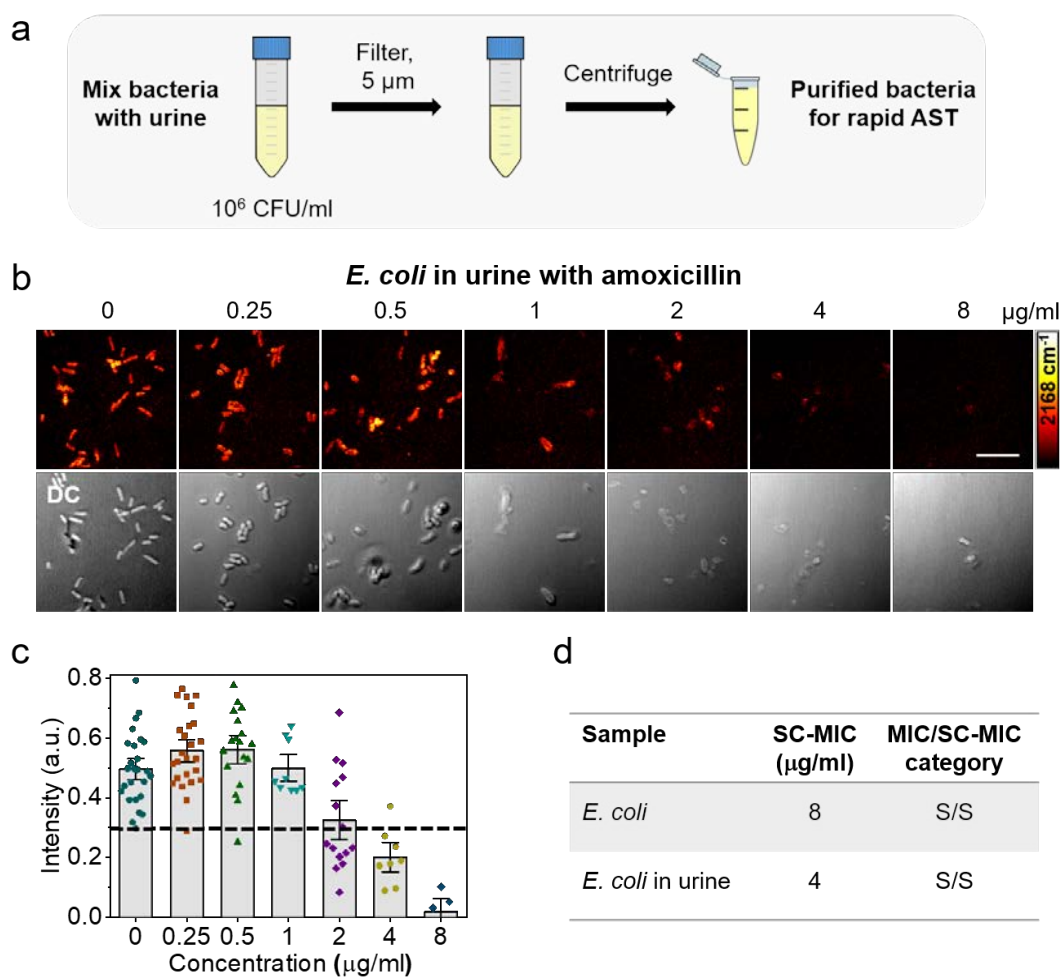

**Figure 4.** SC-MIC determination after 1-h culture of *E. coli* in urine. (a) Bacterial purification protocol for bacteria in urine for rapid AST by SRS imaging of D$_2$O metabolic incorporation. (b) SRS and corresponding transmission images of *E. coli* in urine after 1-h culture in D$_2$O-containing medium with the addition of serially diluted amoxicillin. (c) Statistical analysis of C-D intensity in bacteria in (b). Number of cells per group ≥ 10. (d) Comparison of SC-MIC and susceptibility category for *E. coli* isolate and *E. coli* in urine. Error bars represent SEM. Scale bar: 10 μm.

## 2.6. SC-MIC for bacteria in blood environment

As compared with urine, blood includes a lot of blood cells and presents a much bigger challenge for *in situ* analysis of bacterial activity. To investigate the potential of rapid AST by SRS imaging of D$_2$O metabolic incorporation for clinical bloodstream infections (BSI) samples, we tested *P. aeruginosa* spiked in human blood. Bacteria were first added to blood at a final concentration of ~$10^6$ CFU/ml (**Figure 5a**). Then, water was added to the mixture to lyse the blood cells. After filtration and centrifugation, the purified bacterial samples were used for SC-MIC measurements. The whole procedure for sample preparation took about 15 min. After culture in $D_2O$ medium, SRS images at the C-H vibration showed a lot of debris or blood cells still left in the purified bacterial samples (**Figure 5b**). While, at the same area, the SRS image of C-D vibration was dominatated by bacterial signal. The reason is that deris or red blood cells do not have metabolic activity to incorporate $D_2O$ unlike live bacteria. The weak background mostly comes from the cross-phase modulation or photothermal signal of interferent species, which does not affect the quantification of SC-MICs. The off-resonance SRS images further confirmed that the signals in bacteria largely came from the C-D vibration (**Figure 5b**). The SC-MIC value for *P. aeruginosa* in blood after 1-h culture was determined to be 2 μg/ml (**Figure 5c-d**), which agreed with the SC-MIC or MIC for *P. aeruginosa* in growth medium (**Figure 5e**). These results showed that SRS imaging of $D_2O$ metabolic incorporation can rapidly determine SC-MIC for bacteria in blood environment.

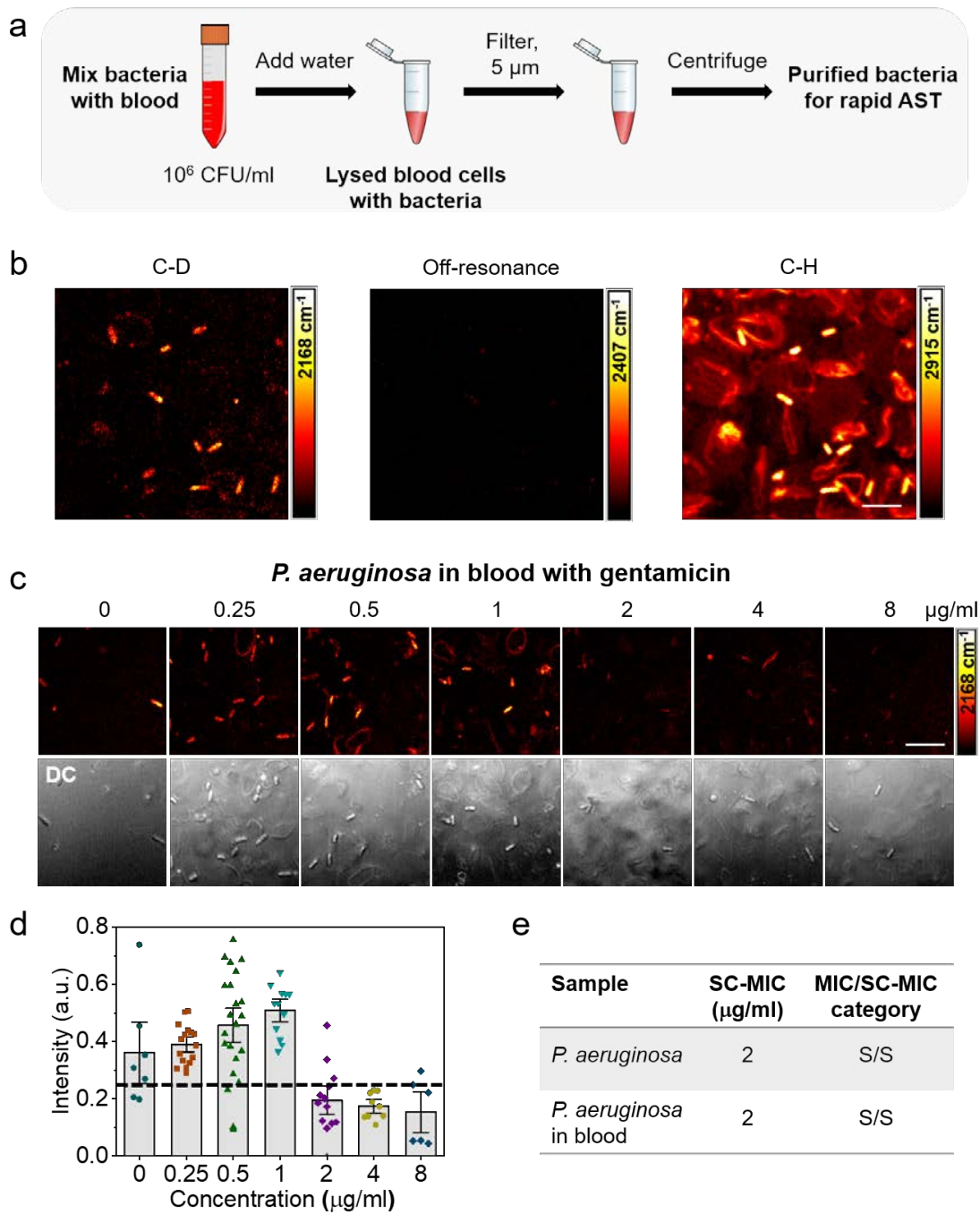

**Figure 5.** SC-MIC determination after 1-h culture of *P. aeruginosa* in blood. (a) Bacterial purification protocol for bacteria in blood for rapid AST by SRS imaging of $D_2O$ metabolic incorporation. (b) SRS images at C-D, off-resonance (2407 cm$^{-1}$), and C-H of bacteria in blood after 1-h culture in $D_2O$ containing medium. (c) SRS and corresponding transmission images of *P. aeruginosa* in blood after 1-h culture in $D_2O$-containing medium with the addition of serially diluted gentamicin. (d) Statistical analysis of C-D intensity in bacteria in (c). (e) Comparison of SC-MIC and susceptibility category for *P. aeruginosa* isolate and *P. aeruginosa* in blood. Error bars represent SEM. Scale bar: 10 μm.

We note that bacterial concentration in the spiked urine and blood samples was $10^6$ CFU/ml in our tests (**Figure 4a** and **Figure 5a**). Since clinically positive UTI samples usually contain more than $10^5$ CFU/ml of bacteria, the bacterial concentration can easily reach $10^6$ CFU/ml or higher after centrifugation.[8b] The bacterial concentrations in positive blood cultures range from $10^6$ to $10^9$ CFU/ml.[23] Therefore, our SC-MIC measurement can be directly used for UTI or positive blood culture samples.

## 3. Discussion

The current work demonstrates a rapid, high-throughput platform that can determine the susceptibility of bacteria in MHB medium, urine and blood by SRS imaging of $D_2O$ incorporation at a single bacterium level. Metabolic incorporation of $D_2O$, which is used for biomolecule synthesis, was monitored in a single bacterium by SRS imaging of C-D bonds. Metabolic response was probed in as short as 10 min after culture in $D_2O$ medium. SC-MIC was obtained in less than 2.5 h from colony to results. The SC-MIC results of 37 sets of samples, which included 8 major bacterial species and 14 different antibiotics, were systematically studied and validated by MIC determined by the broth microdilution method, with a category agreement of 94.6% and 5.4% minor error. Furthermore, we investigated the feasibility of our method to study samples in complex biological environments. The SC-MIC can be determined after 1-h culture of bacteria in urine and blood, which is considered a tremendous reduction in analysis time as compared with the conventional broth microdilution method.

Previously, we monitored the metabolic incorporation of glucose-$d_7$ in isolated bacteria or fungi using SRS microscopy.[19, 24] Though glucose is the preferred carbon source for most bacterial growth,[25] glucose-$d_7$ itself contains C-D bonds which cause a large background in the SRS image. In contrast, cellular metabolic incorporation of $D_2O$ depends on chemical transformation by forming C-D bonds in newly synthesized biomolecules, such as lipids, proteins and DNA.[20a, 26] In the present study, we monitored the $D_2O$ metabolic incorporation by tracking the speed and amount of C-D bond formation. Significantly, the C-D vibration is spectrally separated from the O-D vibration in $D_2O$, allowing for background-free SRS measurement of bacterial metabolic activity in a complex environment such as whole blood. Another innovation of this study is the use of femtosecond pulses, which significantly increased the signal to noise ratio and the imaging speed accordingly.

It is known that stationary-phase and non-dividing bacteria are common in many persistent infections (e.g., endocarditis and osteomyelitis) and in biofilm-associated infectious diseases (e.g., periodontitis and cystic fibrosis).[27] To evaluate the potential of our SRS metabolic imaging method for non-dividing bacteria, we investigated the metabolic dynamics of $D_2O$ incorporation in *E. coli* starting from different phases, lag, log, and stationary phase (**Figure S8**). Interestingly, we observed similar metabolic dynamics during the same period of time, which is consistent with the growth curves with optical density measurements. Hence, our SRS metabolic imaging measurement can be potentially applied to determine the susceptibility of non-dividing bacteria, which is beyond the reach of conventional culture method.

A few groups reported coherent Raman imaging of $D_2O$ activity inside mammalians. Potma et al. used CARS microscopy to minor $D_2O$ entry into a cell in real time.[28] Shi et al. demonstrated picosecond SRS imaging of $D_2O$ metabolism in mammalian cells after 1-day incubation and in live animals after at least 2-day treatment.[20a] Compared to mammalian cells, imaging $D_2O$ metabolic activity in a micron-sized bacterium is challenging. Here, we deployed a few strategies to achive good signal to noise ratio in a single scan. First, stimulated Raman loss is measured, where most excitation power in on the Stokes beam to minimize photodamage to the specimen. Second, femtosecond pulses are used for excitation of the broad C-D vibrational bands, which improved the signal to noise ratio by 5 times compared to picosecond pulses. Third, the cross phase modulation background is minimized by placing the bacteria on a poly-L-lysine-coated glass substrate and covered with phosphate-buffered saline solution.

Because NADPH is ubiquitously used in cell metabolism, our SRS metabolic imaging method has the potential of being broadly used for rapid AST in various strains and can be extended to determine the susceptibility in fungal infections. Another exciting application of this method is for slowly growing bacteria, like *Mycobacterium tuberculosis* which doubles roughly once per day and has a remarkably slow growth rate.[29]

For clinical specimens, each sample requires hours of pre-incubation to obtain bacterial isolates. Methods based on nanoliter array,[8b] digital nucleic acid quantification,[8a] and Raman spectroscopy[15c] have been developed to perform AST for clinical urine samples. Compared with urinary tract infections, bloodstream infections or sepsis are more life-threatening cases,[8a, 30] where rapid AST is urgently needed. Methods based on microscopic imaging,[31] commercial automatic systems,[32] and mass spectrometry[33] allow for direct AST from positive blood cultures. Yet, these methods rely on bacterial growth and take at least 6 h of incubation. In this work, we demonstrate in situ SRS imaging of $D_2O$ metabolic incorporation in single bacteria at a clinically relevant concentration ($10^5$~$10^6$ CFU/ml) in either urine or whole blood. This capacity paves the way toward clinical translation of our technology. While we need to know the antimicrobial susceptibility of bacteria, MIC determination is even more significant in clinics to avoid excess dosage of antibiotics to patients to cause potential side effects.[34] We would emphasize that our SC-MIC method is capable of detecting MICs and susceptibility classification for each strain/antibiotic. Compared with the spontaneous Raman microscopy, our method requires tremendously reduced data acquisitiom time (ca. 600 times less) to obtain MIC results due to orders-of-magnitude signal enhancement. Based on our method, the MICs are determined after 1-h antibiotics treatment and 30-min mixture of $D_2O$ and antibiotics incubation into bacteria in urine and blood. Each SRS image, containing at least 10 bacteria, was acquired within ca. 1 second in one single shot, while it takes about 10 min by spontaneous Raman measurement. We estimate the total MIC assay time to study 10 antibiotics per strain/antibiotic set is less than 2.5 h from sample to MIC results, which is much more efficient and competitive in determining MICs.

Finally, given the importance of identification of pathogens in clinical decision-making, our SRS metabolic imaging can be integrated with diagnostic methods that are capable of rapid identification of pathogens, for example, matrix-assisted laser desorption ionization-time-of-flight (MALDI-TOF) mass spectrometry.[23, 33b, 35], Integration of these *in situ* analysis tools and translation into clinic could potentially eliminate the "culture to colony" paradigm, thus allowing for on-time identification of correct antimicrobial agents for precise treatment.

## Materials and Methods

**SRS microscope.** A femtosecond (fs) pulsed laser (InSight DeepSee, Spectra-Physics) with an 80-MHz repetition rate and dual outputs was employed for excitation. One 120 fs laser with tunable 680–1100 nm wavelength served as the pump beam. The other 220 fs laser, centered at 1040 nm wavelength and served as the Stokes beam, was modulated by an acousto-optical modulator (AOM, 1205-C, Isomet) at ~2.4 MHz. The two beams were collinearly combined through a dichroic mirror. When spectral focusing is needed for hyperspectral SRS, both beams were chirped with two 15 cm long SF57 glass rods. After chirping, the pulse durations of the pump and Stokes laser were 1.9 ps and 1.3 ps, respectively. For implementation of SRS imaging with femtosecond pulses, the glass rods were removed from the path. The pump and Stokes beams were directed into a lab-built laser scanning microscope with a 2D galvo mirror for laser scanning. A 60× water objective (NA=1.2, UPlanApo/IR, Olympus) was used to focus the lasers to the sample, and an oil condenser (NA=1.4, U-AAC, Olympus) was used to collect the signal from the sample. Two filters (HQ825/150m, Chroma) were used to filter out the Stokes beam, the pump beam was detected by a photodiode (S3994-01, Hamamatsu) and the stimulated Raman signal was extracted by a lock-in amplifier (HF2LI, Zurich Instrument).

To image bacteria at the C-D vibrational frequency, the pump wavelength was tuned to 852 nm, and the power at the sample was ~8 mW; the Stokes power at the sample was ~40 mW. Each image contains 200 × 200 pixels and the pixel dwell time is 30 μs, resulting in an image acquisition time of ~1.2 seconds.

**Bacterial strains.** Bacterial strains used in this study (**Table S2**) were obtained from the Biodefense and Emerging Infections Research Resources Repository (BEI Resources) and the American Type Culture Collection (ATCC).

**Sample preparation.** To make $D_2O$ containing Lauria-Bertani broth (LB) (Sigma Aldrich) or cation-adjusted cation-adjusted Mueller-Hinton Broth (MHB) (Thermo Fisher Scientific) media, $D_2O$ was first mixed with purified water, then LB or MHB powder was added to the solution. The solution was sterilized by filtering. To prepare bacterial samples for SRS imaging, bacterial strains were cultured in normal medium to reach the logarithmic phase, then bacteria were diluted to ~$10^6$ CFU/ml in the $D_2O$-containing medium. After a controlled culture time, 500 μl sample was centrifuged, washed twice with purified water, fixed by 10% (w/v) formalin solution (Thermo Fisher Scientific) and deposited to an agarose gel pad or poly-L-lysine coated coverglasses. To prepare samples for spontaneous Raman spectroscopy, bacteria were cultured in normal or 70% $D_2O$-containing medium for 2 h, and then centrifuged and washed twice with purified water to remove the culture medium. For the time lapse experiment, *P. aeruginosa* ATCC 47085 were cultured in 70% $D_2O$ containing medium for up to 3 h. At different time points, 500 μl bacteria were centrifuged, washed twice with purified water, and deposited on a poly-L-lysine coated slide for imaging.

**$D_2O$ toxicity measurement.** $D_2O$ toxicity measurement was done in 96-well plates. *P. aeruginosa* ATCC 47085, *E. coli* BW25113, or *S. aureus* ATCC6538were cultivated into $D_2O$- containing LB medium with D$_2$O concentrations ranging from 0 to 100%, and incubated at 37 °C. At each time point, optical densities (OD) of the samples were measured at 600 nm wavelength.

**Data analysis.** To get the average C-D signal intensity, the SRS images were opened in ImageJ software (ImageJ, NIH), single bacterial cells were selected, and the average intensity of the selected area was calculated. The same procedure was applied for other bacteria in the same image. For bacteria that have filamentary formation, a section of one filament was selected for calculation of SRS intensity.

**Spontaneous Raman spectroscopy.** Bacteria in solution were deposited on a coverglass for spontaneous Raman measurement. Spontaneous Raman spectra of bacteria were acquired with an inverted Raman spectrometer (LabRAM HR evolution, Horiba scientific) with 532 nm laser source. The laser power at the sample was ~12 mW after a 40× air objective, and the acqusition time was 10 s. The grating was 600 l/mm.

**Broth microdilution method.** Bacteria were cultured in D$_2$O-containing MHB medium in 96-well plates. Antibiotics, using triplicate samples, were added to the plate and two-fold serially diluted. Plates were then, incubated aerobically at 37° C for ~20 h. MIC was categorized as the concentration at which no visible growth of bacteria was observed.

Additional experimental results and data analysis are available in the supplementary information.

**Data availability.** The data that support the plots within this paper and other findings of this study are available from the corresponding author upon reasonable request.

## Acknowledgements

This work was supported by NIH R01AI141439 to J.X.C. and M.S.. We thank Lijia Liang for helping with the MIC measurements.

## Author contributions

J.X.C and W.H. conceived the idea. J.X.C., W.H., M.Z., M.S., and P.W. designed the experiments. M.Z., W.H., N.S.A., J.L., P.T.D. and C.Z. conducted the experiments and analyzed the data. M.Z., W.H., and J.X.C. co-wrote the manuscript. All authors have contributed to discussing and editing the manuscript, and given approval to the final version of the manuscript.

**Corresponding Author:** Correspondence to Ji-Xin Cheng (jxcheng@bu.edu) and Weili Hong (weilihong@buaa.edu.cn).

## Competing interests
The authors declare no competing interests.

# Supplementary Information

# Rapid Determination of Antimicrobial Susceptibility by Stimulated Raman Scattering Imaging of D$_2$O Metabolic Incorporation in a Single Bacterium

Meng Zhang[1], Nader S. Abutaleb[2], Junjie Li[1], Pu-Ting Dong[1], Cheng Zong[1], Pu Wang[3,4], Mohamed N. Seleem[2,5], Weili Hong[3,*], Ji-Xin Cheng[1,6,*]

[1] Department of Electrical & Computer Engineering, Boston University, Boston, Massachusetts 02215, USA.
[2] Department of Comparative Pathobiology, Purdue University, West Lafayette, Indiana 47907, USA.
[3] Beijing Advanced Innovation Center for Biomedical Engineering, Beihang University, Beijing, 100191, China
[4] Vibronix Inc., West Lafayette, Indiana 47906, USA.
[5] Purdue Institute of Inflammation, Immunology, and Infectious Disease, West Lafayette, Indiana 47907, USA.
[6] Department of Biomedical Engineering, Department of Chemistry, Photonics Center, Boston University, Boston, Massachusetts 02215, USA.
*Corresponding authors: jxcheng@bu.edu; weilihong@buaa.edu.cn.

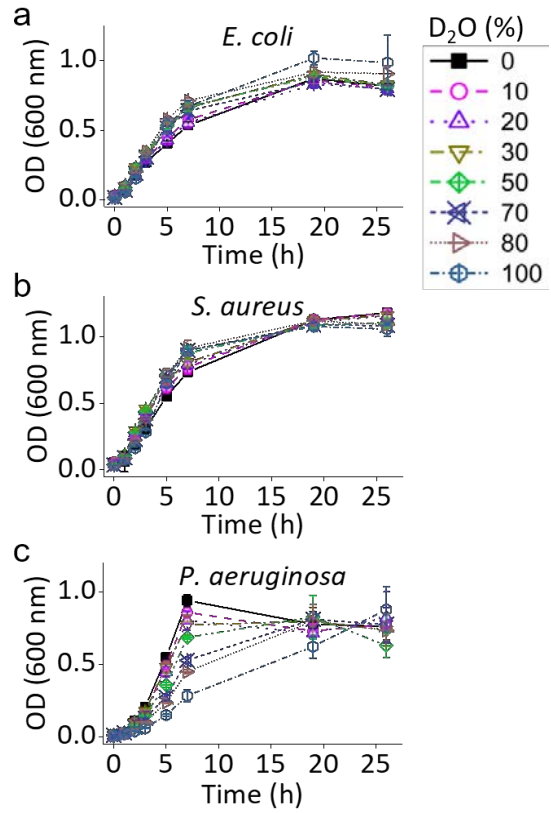

**Figure S1.** Testing D$_2$O toxicity to bacterial growth. Growth curve of *E. coli*. (a), *S. aureus* (b) and *P. aeruginosa* (c) cultured in LB medium with different D$_2$O concentrations. Error bars indicate standard deviation values (number of measurements = 5).

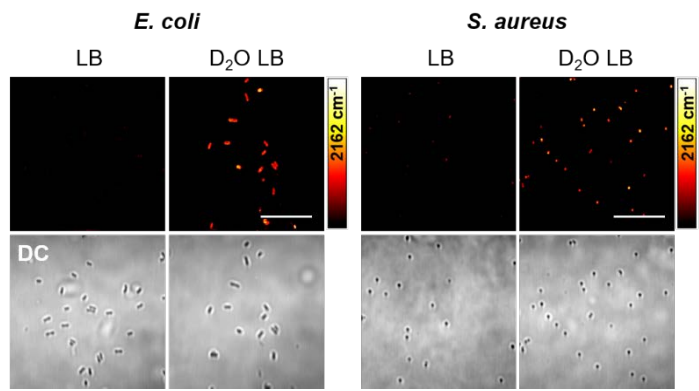

**Figure S2.** SRS and corresponding transmission images of *E. coli* and *S. aureus* after being cultured in LB and D$_2$O containing LB medium for 3 h. Scale bar: 20 μm.

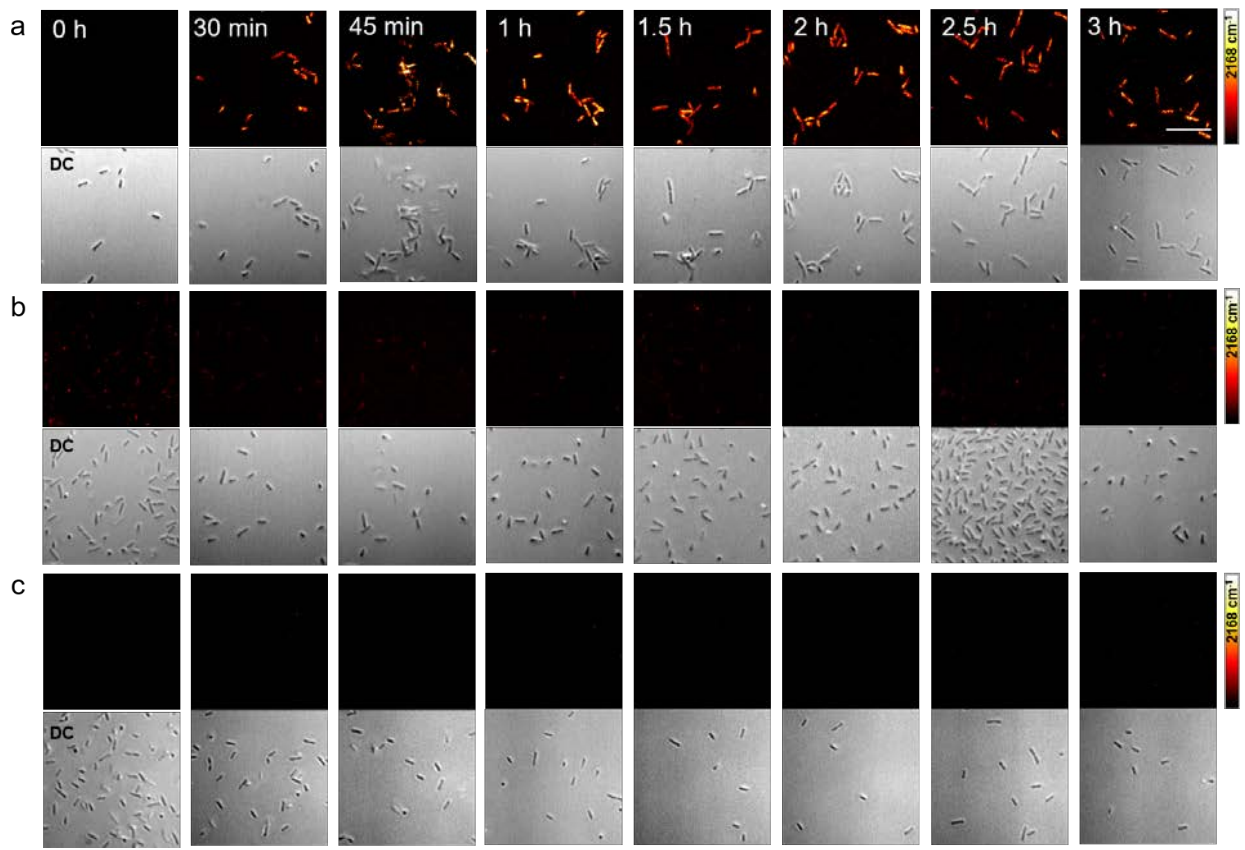

**Figure S3.** SRS imaging of D$_2$O metabolic activity in (a) 37 °C live, (b) 4 °C live and (c) 37 °C fixed *P. aeruginosa* ATCC 47085 cultured in 70% D$_2$O MHB medium with different incubation times. Scale bar: 10 μm.

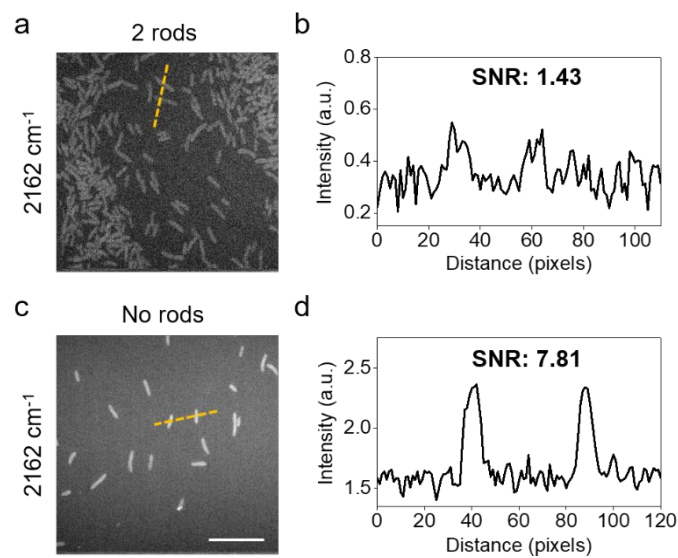

**Figure S4.** Femtosecond SRS improves signal to noise ratio (SNR) at C-D vibrational region over the chirped SRS. (a) SRS image at ~2162 $cm^{-1}$ of *P. aeruginosa* 47085 cultured in 70% $D_2O$ containing LB medium for 30 min with picosecond pulses generated by chirping with two SF57 glass rods. (b) Intensity plot of the orange line over bacteria in (a). (c) SRS image at ~2162 $cm^{-1}$ of *P. aeruginosa* cultured in 70% $D_2O$ containing LB medium for 30 min with femtosecond pulses. (d) Intensity plot of the orange line over bacteria in (c). Scale bar: 20 μm.

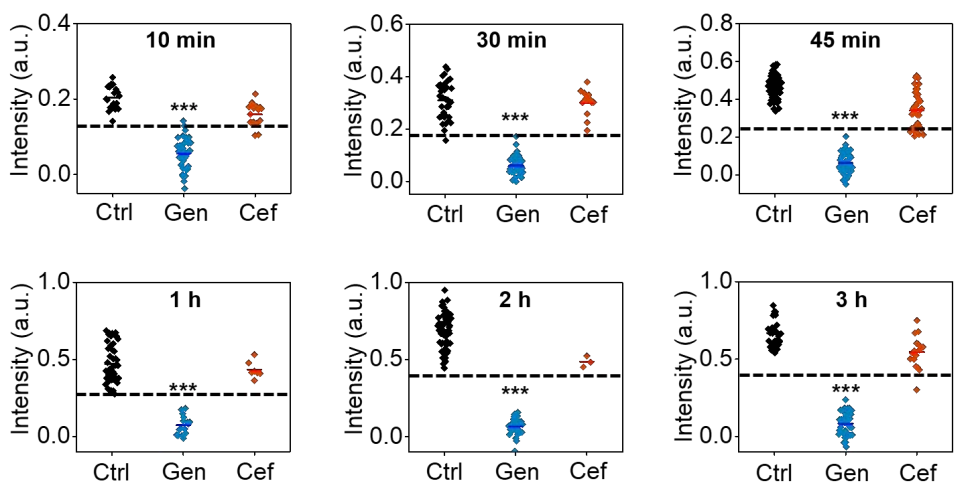

**Figure S5.** Statistical analysis of C-D intensity in *P aeruginosa* without treatment and with gentamicin or cefotaxime treatment for different time. Black bars indicate the threshold that discriminates susceptible (gentamicin) and resistant (cefotaxime) groups.

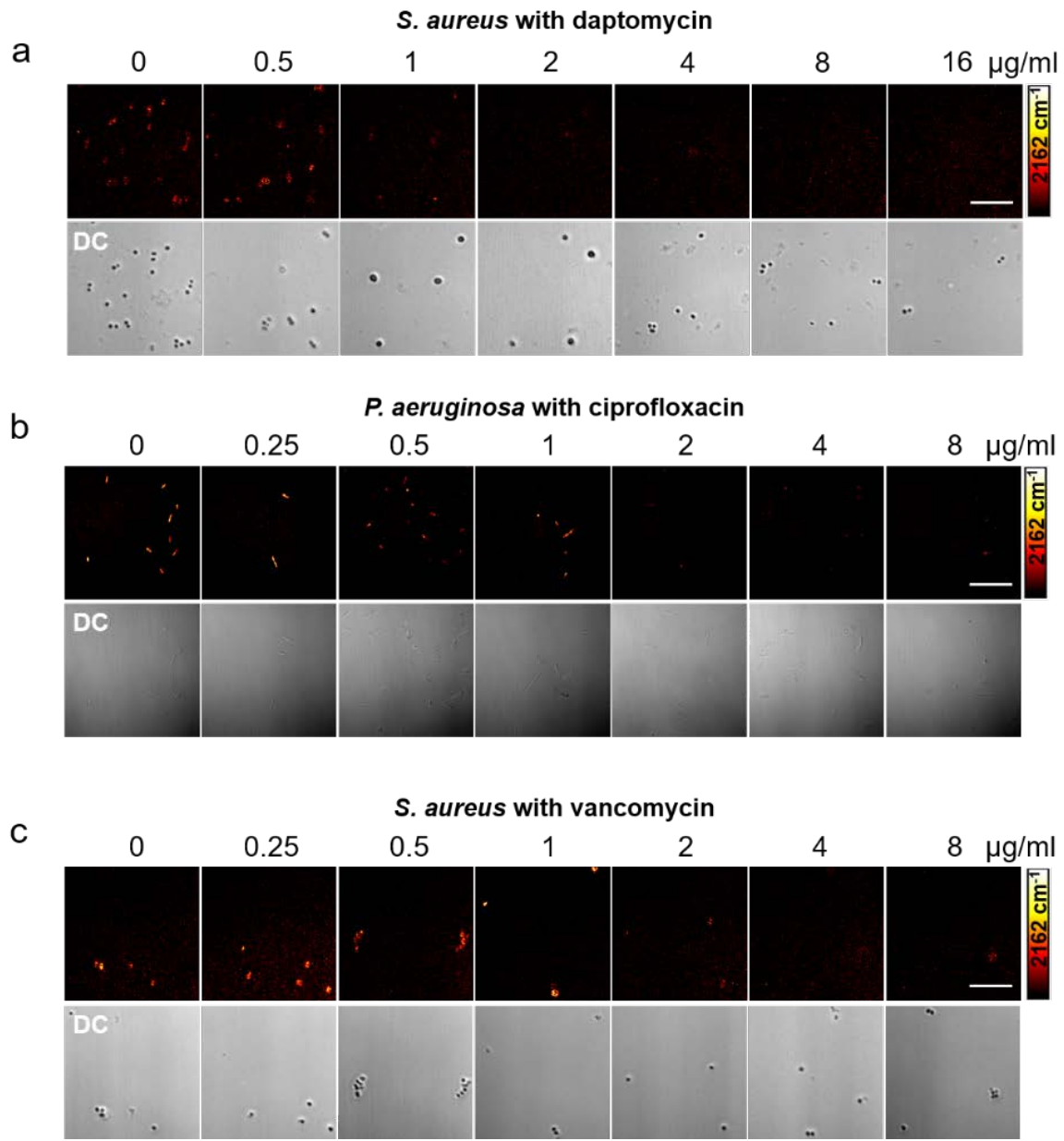

**Figure S6.** SC-MIC determination for *S. aureus* and *P. aeruginosa* treated with different mechanism of action. (a-c) SRS and corresponding images of bacteria cultivated in antibiotics for 30 min and D$_2$O containing medium for another 30 min. Scale bar: (a, c) 10 µm; (b) 20 µm.

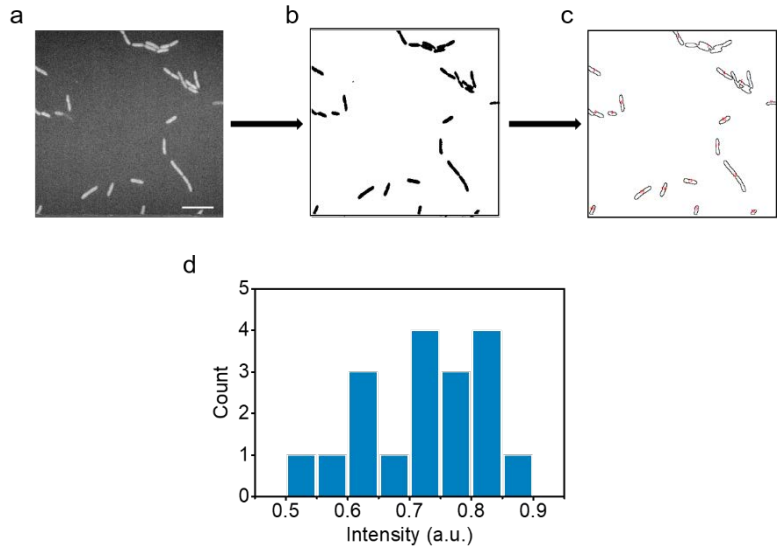

**Figure S7.** Image processing and data interpretation. Scale bar: 10 μm.

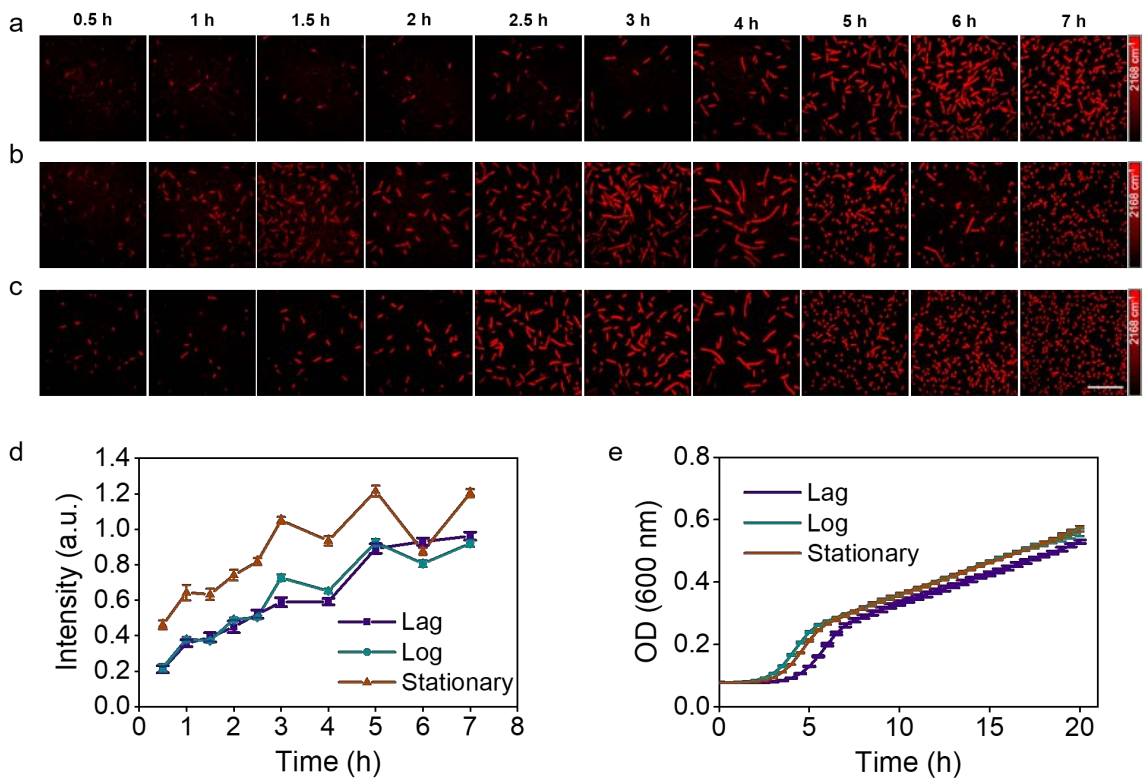

**Figure S8.** SRS imaging of D$_2$O metabolic activity in *E. coli* BW 25113 of (a) lag phase, (b) log phase and (c) stationary phase cultured in 70% D$_2$O MHB medium with different incubation time. Scale bar: 20 μm.

**Table S1.** List of MICs obtained in normal MHB medium and in 70% D$_2$O-containing MHB medium.

| Bacteria strains | Antibiotics | MIC in normal MHB (μg/ml) | MIC in 70% D$_2$O MHB (μg/ml) |
|---|---|---|---|
| S. aureus ATCC 6538 | Vancomycin | 1 | 0.5 |
| | Linezolid | 2 | 1 |
| | Daptomycin | 1 | 0.25 |
| | Gentamicin | 0.5 | 0.25 |
| | Erythromycin | 0.06 | 0.125 |
| | Trimethoprim/ Sulfamethoxazole | 0.1/0.5 | 0.1/0.5 |
| Methicillin-resistant S. aureus (MRSA) NRS 385 (MRSA USA500) | Trimethoprim/ Sulfamethoxazole | 256/1280 | >12.8/64 |
| Methicillin-resistant S. aureus (MRSA) NRS 119 | Linezolid | 64 | >64 |
| Vancomycin-resistant S. aureus (VRSA) NR-46419 (VRSA 9) | Vancomycin | >128 | >64 |
| A. baumannii ATCC BAA-747 | Amikacin | 1 | 0.5 |
| | Ciprofloxacin | 0.25 | 0.06 |
| | Gentamicin | 0.5 | 0.25 |
| | Doxycycline | 0.125 | 0.5 |
| A. baumannii ATCC BAA 1605 | Gentamicin | >512 | 64 |
| | Doxycycline | 8 | >64 |
| | Ciprofloxacin | 128 | 64 |
| K. pneumoniae ATCC BAA 1706 | Tobramycin | 32 | 32 |
| | Ciprofloxacin | 8 | 2 |
| | Imipenem | 2 | 1 |
| K. pneumoniae ATCC BAA 1705 | Imipenem | 16 | 16 |
| E. cloacae ATCC BAA 1143 | Gentamicin | 0.5 | 0.5 |
| | Cefotaxime | >128 | >64 |
| S. enterica ATCC 700720 | Amoxicillin | 1 | 1 |
| | Tobramycin | 1 | 1 |
| E. coli ATCC 25922 | Cefotaxime | 0.06 | 0.06 |
| | Tobramycin | 1 | 0.5 |
| E. coli BW 25113 | Tobramycin | 0.5 | 1 |
| | Amoxicillin | 8 | 4 |
| P. aeruginosa 1133 | Colistin | >512 | 32 |
| P. aeruginosa ATCC 47085 (PAO1) | Cefotaxime | 16 | 8 |
| | Colistin | 2 | 0.5 |
| | Gentamicin | 4 | 1 |
| | Amikacin | 2 | 2 |
| | Tobramycin | 0.5 | 0.5 |
| | Ciprofloxacin | 0.125 | 0.125 |
| E. faecalis NR-31970 | Daptomycin | 4 | 0.06 |
| E. faecalis HM-335 | Daptomycin | 32 | 32 |

**Table S2.** List of bacterial strains used in this study.

| Bacterial strains | Source/ Description |
|---|---|
| *S. aureus* ATCC 6538 | Quality control strain<br>Isolated from human lesion |
| Methicillin-resistant *S. aureus* (MRSA) NRS 385 (MRSA USA500) | Isolated from a bloodstream sample in Connecticut, USA. It is a hospital-acquired methicillin-resistant S. aureus (HA-MRSA) strain. |
| Methicillin-resistant *S. aureus* (MRSA) NRS 119 | Isolated in 2001 from an 85-year-old male with dialysis-associated peritonitis in Massachusetts, USA. |
| Vancomycin-resistant *S. aureus* (VRSA) NR-46419 (VRSA 9) | Isolated in 2007 in Michigan, USA from a left plantar foot wound of a 54-year-old female who recently received a 4-week course of vancomycin and levofloxacin to treat osteomyelitis of the left metatarsals |
| *A. baumannii* ATCC BAA-747 | Human clinical specimen<br>Isolated from ear pus<br>Quality control strain |
| *A. baumannii* ATCC BAA 1605 | Isolated from sputum of military personnel returning from Afghanistan entering a Canadian hospital, June 30, 2006<br>Resistant to Ceftazidime, Gentamicin, Ticarcillin, Piperacillin, Aztreonam, Cefepime, Ciprofloxacin, Imipenem, and Meropemem.<br>Sensitive to Amikacin and Tobramycin |
| *K. pneumoniae* ATCC BAA 1706 | Quality control strain |
| *K. pneumoniae* ATCC BAA 1705 | Isolated from a urine sample of 42-year-old male<br>*blaKPC* positive<br>Carbapenem-resistant (Imipenem and Ertrapenem) |
| *E. cloacae* ATCC BAA 1143 | A control strain, derived from an existing strain, 1982 |
| *S. enterica* ATCC 700720 | Wild-type strain isolated from a natural source, 1948<br>Used for: Emerging infectious disease research |
| *E. coli* ATCC 25922 | Quality control strain for antibiotics susceptibility testing |
| *E. coli* BW 25113 | The parent strain of the Keio collection comprising nearly 4,000 single-gene deletion mutants |
| *P. aeruginosa* 1133 | Quality control strain<br>Polymexin B- and colistin-resistant |
| *P. aeruginosa* ATCC 47085 (PAO1) | A laboratory strain constructed by stable integration of a mini-D3112 transposable element containing sequences allowing lacZalpha complementation<br>Used for transformation host and opportunistic pathogen research |
| *E. faecalis* NR-31970 | Isolated in 2001 from a urine sample in Michigan, USA.<br>Resistant to erythromycin and gentamicin |
| *E. faecalis* HM-335 | Isolated in 2004 from the blood of a 64-year-old female hemodialysis patient with fatal bacteremia after treatment with daptomycin.<br>Resistant to daptomycin |